\documentclass[12pt]{article}
\newcommand{\bm}[1]{\mbox{\boldmath $#1$}}

\newcommand{\xrm}[1]{{\textstyle \mbox{\rm #1}}}
\newcommand{\x}[1]{{\textstyle #1}}

\def\o782{\mbox{$\omega$(782)}}
\begin{document} 
\title{\bf The pionic width of the \bm{\omega(782)} meson within\\
a well-defined, unitary quantum field theory\\
of (anti-)particles and (anti-)holes\\ [.5cm]}
\author{{\bf\Large Frieder Kleefeld}$^{\, a}\!$
\footnote{{\tt kleefeld@cfif.ist.utl.pt} (corresponding author)}$\;$,
{\bf\Large Eef van Beveren}$^{\, b}\!$
\footnote{\tt eef@teor.fis.uc.pt}
$\;${\bf\Large and George Rupp}$^{\, a}\!$
\footnote{\tt george@ajax.ist.utl.pt}
\\ [.5cm]
$^{a}${\normalsize\it Centro de F\'{\i}sica das Interac\c{c}\~{o}es
Fundamentais (CFIF),}\\
{\normalsize\it Instituto Superior T\'{e}cnico (IST), Edif\'{\i}cio
Ci\^{e}ncia, Av.\ Rovisco Pais,}\\
{\normalsize\it P-1049-001 Lisboa, Portugal}\\ [.3cm]
$^{b}${\normalsize\it Departamento de F\'{\i}sica, Universidade de Coimbra,}\\
{\normalsize\it P-3004-516 Coimbra, Portugal}\\ [.3cm]
[.3cm]
{\small hep-ph/0101247}
}
\date{\today}
\maketitle

\begin{abstract}
We investigate the indirect generation of the partial width of the
\o782 meson with respect to the reaction
$\o782\rightarrow  \rho(770)\,\pi \rightarrow\pi^+\pi^-\pi^0$
within the context of a ``Unitary Effective Resonance Model'' (UERM),
which has previously been applied to resonant fermionic field
operators, here extended to scalar, pseudoscalar and vector bosonic fields.
The \o782 meson is described by a (quasi-)real field, while the
intermediate $\rho(770)$ meson is considered to be a resonant degree
of freedom, which can be treated consistently within the UERM.
  
In the limit of an infinitesimal width (imaginary part of self-energy),
the UERM yields a consistent
treatment of relativistic quantum field theory (QFT).
Some aspects of the UERM of (anti-)particles and (anti-)holes are
discussed.
\end{abstract}
\vspace{1cm}

\clearpage

\section{Introduction}
Pions are fundamental particles of effective formulations for the
description of strong interactions,
not only because they are the first discovered mediators of nuclear forces,
but also since most hadrons created in experiment may decay either directly,
or through a cascade of decay processes, into pions, which subsequently
evaporate into photons, neutrinos, and leptons. In this article we will
study the decay process of the \o782 meson.
  
In the unitarised meson model (NUMM) of Ref.~\cite{Bev83}
it was found that, among the various two-meson channels which couple
to the non-strange quark+anti-quark system with the \o782 quantum numbers, the
$\rho$(770)$\pi$ channel is the main responsible for the correct
central mass position of the \o782 meson. This model observation does
not stand alone, but was already suggested as early as in 1962 by
M.~Gell-Mann, D.~Sharp, and W.G.~Wagner \cite{GSW}, and has been studied
in a number of publications \cite{LNSV,rhompi} since.
It is moreover supported by experiment, as we will explain below.

In the NUMM one considers implicit mixing of the non-strange and strange
$q\bar{q}$ sectors through their common $K\bar{K}$, $K^{\ast}\bar{K}$,
$K\bar{K}^{\ast}$, and $K^{\ast}\bar{K}^{\ast}$ channels.
The resulting system simultaneously describes both the \o782 and $\phi$(1020)
mesonic states, and their radial excitations.
As a consequence, the NUMM $\phi$(1020) mesonic state contains a small
non-strange $q\bar{q}$ component, \mbox{which fully} mimics the three-pion
decay width through its $\rho$(770)$\pi$ partial decay width.
From experiment one deduces that at least 80 percent of the three-pion
decay mode of the $\phi$(1020) meson stems from a cascade process through
$\rho$(770)$\pi$ \cite{PP,Par76}.
Hence, when in the NUMM the non-strange component of the $\phi$(1020) meson
cascades dominantly through $\rho$(770)$\pi$ into three pions, then one
may safely assume that the non-strange component of its partner, the \o782
meson, also does.
  
However, with $\rho$ and $\pi$ masses of about 760 MeV and 140
MeV, respectively, in the NUMM the $\rho$(770)$\pi$ channel is closed for
the \o782 meson. Consequently, in the NUMM one ends up with a bound state
(or infinitely sharp resonance) for the model's $\omega$ meson.
There are two ways out of that situation: either one assumes the
existence of many $\rho$ mesons, for instance with a Breit-Wigner-like
probability distribution around the central resonance position \cite{Mui},
or one takes a complex value for the $\rho$ mass.
Both cases evidently describe the cascade process
\begin{equation}
\omega\;\longrightarrow\;\rho\pi\;\longrightarrow\; (2\pi )\pi
\;\;\; .
\label{cascade}
\end{equation}
Substitution in the NUMM of the $\rho$(770)$\pi$ channel by a variety of
$\rho\pi$ channels, each with a different, but real, $\rho$-meson mass,
obviously leads to the numerical difficulty of needing to handle very many
coupled channels and, moreover, to the rather ad hoc choice of their
probability distribution, whereas opting for a complex $\rho$-meson mass
gives rise to non-unitarity.
\clearpage

The reason for the present study is that the $\omega$ width is a clean
application and test of a recently developed technique which allows for
complex masses without violating unitarity.
This formalism, the Unitary Effective Resonance Model (UERM), has been
developed for fermionic degrees of freedom in order to consistently
treat the dynamics of baryonic resonances \cite{Kle99}.
  
In this paper the UERM will be extended to bosonic systems.
The resulting model \mbox{will then} be applied to the description of the
effective two-pion system with the $\rho$-meson quantum numbers.

The organisation of this paper is as follows.
In section~\ref{UERM} the concept of the UERM is extended to a
well-defined quantum field theory (QFT) of scalar, pseudoscalar and
vector bosonic
(anti-) particles and (anti-)holes with infinitesimal or non-zero width.
Several interpretational points of this QFT are addressed.
In section~\ref{UERMresult} we calculate the UERM result for the cascade
process (\ref{cascade}), which consistently
demonstrates the indirect generation of the partial decay width of a
``real'' $\omega$ meson into three pions via the formation of an
intermediate resonant $\rho$ meson.
The conclusions and a discussion are presented in section~\ref{conclusion}.
\clearpage

\section{The UERM extension to bosons}
\label{UERM}
The present theoretical understanding of particle dynamics at low and
intermediate energies is formulated in terms of
effective field theories (EFTs), i.e., QFTs
of {\it effective} \/degrees of freedom, which are assumed to be
the low-energy limit of underlying field theories describing the dynamics
of {\it elementary} \/(point-like) degrees of freedom.
Some of the effective degrees of freedom may be considered stable, like
pions in strong interactions, others are experimentally known to have
finite lifetimes, like the $\rho$ meson, which shows they are complex
dynamical systems under time evolution.
In this paper we address the question of how to incorporate the latter
effective degrees of freedom in a consistent way in the EFT for mesons.
For effective fermionic degrees of freedom with finite lifetimes,
like baryonic resonances, the formalism has been extensively described in
Ref.~\cite{Kle99}.
Here we study the extension to effective bosonic degrees of freedom,
leading to a well-defined QFT of particles, anti-particles, holes,
and anti-holes.

The UERM is not only well-defined with respect to boundary
conditions and unitarity, it reproduces the results obtained within the
commonly used QFTs in the fermionic sector exactly, while yielding small,
yet non-trivial corrections in the bosonic sector stemming from the
negative-energy states of bosons here defined to be bosonic holes and
anti-holes. Moreover, it puts
the field-theoretic treatment of bosonic degrees of freedom on the same
footing as that of the fermions,
it provides a new interpretation of
the meaning of anti-particles and holes, and it allows a field-theoretic
treatment of effective degrees of freedom with a non-zero
width. The concept of a Dirac sea is no longer necessary and the
Klein-Gordon equation is allowed to have negative-energy solu\-tions
without immediate Bose-Einstein condensation of bosonic states to
infinite negative energy.

Furthermore, there is a close correspondence between Fock state vectors
in UERM and (anti-) Gamow states in quantum mechanics (QM).
Here, the Fock state vectors of (anti-)particles in UERM correspond to
Gamow states in QM, whereas the Fock state vectors of (anti-)holes in UERM
correspond to anti-Gamow states in QM.
As a consequence several properties of states with non-zero or infinitesimal
width can also be observed and studied in quantum-mechanical
models treating resonances on the basis of (anti-)Gamow states 
\cite{Ful69,Boh81}.
The presence of anti-Gamow states in scattering theory is essential in order
to consistently handle time-reversal invariance
and to restore unitarity, like the existence of
(anti-)holes in UERM, which can be studied in more detail using separable
potentials \cite{Ful69}.
\clearpage

\subsection{(Anti-)particles and (anti-)holes}
\label{qft1}
\subsubsection{Introduction}
Consider the generating functional of a real free bosonic field $\phi (x)$
with the free action $S_0\,[\phi\,]$ in Minkowski space
(metric tensor $g_{\,\mu\nu} =$ diag$(+,-,-,-)$):
\begin{eqnarray} Z_0  & = & \int D[\phi ] \; \exp \Big( \x{i \, S_0\,
[\phi ]} \Big) \nonumber \\
 & & \nonumber \\
 & = & \int D[\phi ]
\; \exp \left( i \int d^{\,4}x \; \frac{1}{2} \left( (\partial_\mu \,
\phi\, (x) ) (\partial^{\,\mu} \,\phi \,(x) )  - \, m^2 \,
\phi^2 (x) \right) \right)\; .
\end{eqnarray}
This generating functional is only a well-defined convergent Gaussian
path integral, if the (real) mass $m$ of the boson is analytically continued
into the complex plane by introducing an infinitesimal negative imaginary
part, to yield an exponential fall off of the integrand. Hence, one has to
perform the replacement $m\rightarrow M = m - i \, \varepsilon$ with $
\varepsilon$ being real, positive, and infinitesimally small.

A consequence of this substitution becomes more evident if we allow
for a finite width $\Gamma /2$, i.e., $M = m - i \, \Gamma /2$.
The action (and therefore also the effective action) and the
Hamilton operator are no longer hermitian, which yields a non-unitary
S matrix or Dyson operator. The eigenstates of the Hamilton operator
(or the Fock states) are no longer orthogonal, and, simply speaking,
the ``bras'' ($<\ldots|$) are no longer hermitian conjugates of the
``kets'' ($|\ldots >$). Of course, the same arguments also apply to a
free fermionic field.
The, at least infinitesimal, non-hermiticity of the
Hamilton operator is indicating that the Fock space in QFT, or the
Hilbert space in QM, is incomplete. This problem has been solved
for charged fermionic resonance fields of finite width within the UERM 
\cite{Kle99} by consistently introducing additional
fields in the Lagrangian in order to restore the unitarity of the theory.

In the same way as in the fermionic case, the real bosonic field $\phi(x)$
has to be replaced by a ``quasi-real'' field $\phi\, (x)$ and its
hermitian conjugate $\phi^+ (x)$. In order to include isospin
at this stage, i.e., charged fields, we start with $N$
``quasi-real'' fields $\phi_r (x)$ ($r = 1,\ldots,N$) of equal complex
mass $M = m - i \, \Gamma / 2\,$, yielding the following free Lagrangian
for what we traditionally call $N$ ``real'' (i.e. uncharged) bosons:
\begin{eqnarray} L^{\,0}_{\,\phi} (x) & = & \sum\limits_{r} \;
\frac{1}{2} \left( (\partial_\mu \,\phi_r (x) ) (\partial^{\,\mu} \,
\phi_r (x) )  - \, M^2 \,
\phi_r (x) \, \phi_r (x) \right) \; + \nonumber \\
 & + & \sum\limits_{r} \;
\frac{1}{2} \left( (\partial_\mu \,\phi_r^+ (x) ) (\partial^{\,\mu} \,
\phi_r^+ (x) )  - \, M^{\ast \, 2} \,
\phi_r^+ (x)\, \phi_r^+ (x) \right) \label{gleich1}
\; .
\label{Lag0}
\end{eqnarray}
The term in the first line of formula (\ref{Lag0})
is well known. It is reponsible for the description
of particles and anti-particles, while the term in the second line is
absent in traditional QFTs. This term is describing what we will
call holes and anti-holes. As we will see later on, {\em with respect to
movement in the forward time direction}, the real part of the energy of
{\em particles and anti-particles} \/is {positive}, whereas the real part
of the energy of {\em holes and anti-holes} \/is {negative}.
On the other hand, {\em with respect to movement in the backward time
direction}, the
real part of the energy of {\em particles and anti-particles} \/is
{negative}, whereas the real part of the energy of {\em holes and
anti-holes} \/is {positve}.
So {\em with respect to movement in the forward time direction},
particles and anti-particles {\em minimise} \/their energy, whereas
holes and anti-holes {\em maximise} \/their energy.
{\em With respect to movement in the backward time direction}, particles
and anti-particles {\em maximise} \/their energy, whereas holes and
anti-holes {\em minimise} \/their energy.
In our language, anti-particles are objects with opposite additive
quantum numbers (e.g.\ charge, parity, $\ldots$) as compared to the respective
particles, except for the energy. Anti-holes have opposite additive
quantum numbers as compared to the respective holes, except for the energy.
This is why, with respect to the forward time direction,
the energy of annihilation radiation of a particle
and an anti-particle is positive and larger than or equal to twice
the mass of the particle, whereas the energy of the annihilation radiation
of a hole and an anti-hole is less than or equal to twice the (negative)
mass of the hole.
So, in the UERM there no longer exists an identification of
anti-particles and holes.

The distinction between anti-particles and holes (or particles and
anti-holes) seems to be quite arbitrary for objects with
inifinitesimal width, yet for the field-theoretic description of
systems of non-zero width it is crucial.

\subsubsection{Canonical quantisation of ``quasi-real'' boson fields}
The classical equations of motion are obtained in the standard manner
by varying
the action $S = \int d^{\,4}x \, L^{\,0}_{\,\phi} (x)$ with respect
to the fields $\phi_r (x)$ and $\phi_r^+ (x)$ ($r = 1,\ldots,N$):
\begin{equation}
(\,\partial^2 + M^2) \,\phi_r (x) \; = \; 0 \quad , \quad (\,\partial^2
+ M^{\ast \, 2}) \,\phi_r^+ (x) \; = \; 0\; .
\label{eqmot}
\end{equation}
Obviously, equations (\ref{eqmot}) are pairwise complex conjugate
to each other, i.e., pairwise equivalent, while each equation contains
two coupled equations for the real and imaginary part of $\phi_r (x)$ and
$\phi_r^+(x)$, respectively.
The standard canonical conjugate momenta to the fields $\phi_r (x)$ and
$\phi_r^+ (x)$ are ($r = 1,\ldots,N$)
\begin{equation}
\Pi_r (x)  = \frac{\delta \,L^{\,0}_{\,
\phi} (x)}{\delta \, (\partial_0 \,\phi_r (x))}  =  
\partial_0 \,\phi_r (x) \; , \;\; \Pi_r^+ (x) = \frac{\delta
\,L^{\,0}_{\,\phi} (x)}{\delta \, (\partial_0 \,\phi_r^+ (x))} 
  = \partial_0 \,\phi_r^+ (x)
\; .\end{equation}
Canonical quantisation in configuration space yields
the {\em non-vanishing} \/equal-time commutation relations
($r,s = 1,\ldots,N$)
\begin{eqnarray}
{ [ \, \phi_r (\vec{x},t) , \Pi_s (\vec{y},t) \, ] }
 & = &
 i\, \delta^{\, 3} (\vec{x} - \vec{y}\,) \; \delta_{rs} \nonumber \\
 & & \nonumber \\
{ [ \, \phi^+_r (\vec{x},t) , \Pi^+_s (\vec{y},t) \, ] }
 & = &
 i\, \delta^{\, 3} (\vec{x} - \vec{y}\,) \; \delta_{rs} \; .
\end{eqnarray}
The classical equations of motion (\ref{eqmot}) can be solved by a standard
Laplace transform.
The corresponding transformation for the
field operators reads ($\omega \, (| \vec{k}\,|) = \sqrt{|\vec{k}\,|^2 +
M^2}\;$ with Re$\,(\omega \, (|\vec{k}\,|)) \ge 0$, $k^{\,\mu} 
 =
(\omega \, (| \vec{k}\,|) , \, \vec{k} \,\, )$, and $k^{\ast\,\mu} 
 =
(\omega^\ast (| \vec{k}\,|) , \, \vec{k} \,\, )\,$)($r = 1,\ldots,N$)
\begin{eqnarray}
\phi_r (x) & = & \int \! \!
\frac{d^3 k}{(2\pi )^3 \; 2\, \omega \, (| \vec{k}\,|)}
\;\;
\Big[ \,
 a \, (\vec{k} , r ) \; e^{\displaystyle - \, i k x} \; + \;
 c^+ \, (\vec{k} , r ) \; e^{\displaystyle i k x}
\,\Big] \nonumber \\
 & & \nonumber \\
\phi_r^+ (x) & = & \int \! \!
\frac{d^3 k}{(2\pi )^3 \; 2\,\omega^\ast (|\vec{k}\,|)}
\;\;
\Big[\,
  c \, (\vec{k} , r ) \; e^{\displaystyle - \, i k^\ast x} \; + \;
 a^+ \, (\vec{k} , r ) \; e^{\displaystyle i k^\ast x}
\,\Big]
\; .
\end{eqnarray}
The consistent {\em non-vanishing} \/commutation relations for creation
and annihilation operators in momentum space are ($r,s = 1,\ldots,N$)
\begin{eqnarray}
{ [ \, a \, (\vec{k},r) \; , \; c^+ (\vec{k}^{\,\prime},s) \, ] }
 & = &
 (2\pi)^3 \, 2 \, \omega \,(|\vec{k}\,|)\; \delta^{\, 3} (\vec{k} - 
\vec{k}^{\,\prime}\,) \; \delta_{rs} \nonumber \\
 & & \nonumber \\
{ [ \, c \, (\vec{k},r) \; , \; a^+ (\vec{k}^{\,\prime},s) \, ] }
 & = &
 (2\pi)^3 \, 2 \, \omega^\ast(|\vec{k}\,|)\; \delta^{\, 3} (\vec{k} - 
\vec{k}^{\,\prime}\,) \; \delta_{rs}
\; .
\end{eqnarray}
At this point it can been seen that creation and annihilation
operators are not hermitian conjugate to one another.
  
Before making further interpretations of creation and annihilation
operators, it is useful to construct the free Hamilton operator in the
standard manner, i.e.,
\begin{eqnarray}
H^{\,0}_{\phi} 
 & = & \int\! d^{\,3}x  \left[ \sum\limits_{r}
\left( \Pi_r (x) \; (\partial_0 \,\phi_r (x))
+  (\partial_0 \,\phi^+_r (x)) \; \Pi^+_r (x) \,
\right) - \, L^{\,0}_{\phi} (x) \right] \nonumber \\
 & & \nonumber \\
 & & \nonumber \\
 & = &
\sum\limits_{r} \;\int\! d^{\,3}k \,\;
\frac{1}{2}\; \omega \,(|\vec{k}\,|)\; \Big(
c^+ (\vec{k},r) \; a (\vec{k},r) +
a\, (\vec{k},r) \; c^+ (\vec{k},r) \; \Big) + \; \nonumber \\
 & + & \sum\limits_{r} \,\;\int\! d^{\,3}k \;
\frac{1}{2}\; \omega^\ast (|\vec{k}\,|)\; \Big(
a^+ (\vec{k},r) \; c\, (\vec{k},r) +
c \, (\vec{k},r) \; a^+ (\vec{k},r) \; \Big)\; .
\label{Neqone}
\end{eqnarray}
Keeping in mind that the the words ``creation'' and ``annihilation'' are
associated with the forward time direction, we can read off from formula
(\ref{Neqone}) the simple identifications
\begin{eqnarray}
c^+\quad  & \leftrightarrow & \quad \mbox{creation operator of
(uncharged) bosonic particle}, \nonumber \\
a^+\quad  & \leftrightarrow & \quad \mbox{creation operator of
(uncharged) bosonic hole}, \nonumber \\
c\;\; \quad & \leftrightarrow & \quad \mbox{annihilation operator of
(uncharged) bosonic hole}, \nonumber \\
a\;\; \quad & \leftrightarrow & \quad \mbox{annihilation operator of
(uncharged) bosonic particle}\; .
\end{eqnarray}
As the (anti-)particle and (anti-)hole
subsectors of the Hamilton operator are not hermitian, the left and right
eigenvectors of these parts of the Hamiltonian are not just related by
hermitian conjugation.
Hence, the Hamilton operator $H^{\,0}_{\phi}$ is ``diagonal'' in a
more generalised sense.
For example, the ``diagonal'' matrix elements of $H^{\,0}_{\phi}$
corresponding to one-(anti-)particle or \mbox{one-(anti-)} hole
states, are  \mbox{$<\!0|\,a \, H^{\,0}_{\phi} \, c^+|0>$} and 
$<\!0|\,c \, H^{\,0}_{\phi} \, a^+|0>$, while the vanishing ``off-diagonal''
matrix elements are $<\! 0|\,c \, H^{\,0}_{\phi} \, c^+|0>$ and $<\!0|\,a
\, H^{\,0}_{\phi} \, a^+|0>$.
A similiar discussion applies 
in general 
to the calculation of expectation values.
 
It is straightforward to evaluate the
following identity for the Feynman propagator of an uncharged bosonic
field  ($r,s = 1,\ldots,N$):
\begin{eqnarray} i \, \Delta_F (x-y) \;\;  \delta_{rs} \;& = &\; 
<0|\,T\,(\,\phi_r\,(x)\, \phi_s\, (y)\,)\,|0>  
 \; = \; i \int\!\frac{d^{\,4}k}{(2\,\pi)^4}\; \frac{e^{-\,i\,k
(x-y)}}{k^2 - M^2} \;\; \delta_{rs}\, . \nonumber \\
\end{eqnarray}

\subsubsection{Canonical quantisation of the ``complex'' boson field}
In order to proceed in the standard manner, we also consider the
simplest case of a charged bosonic field, i.e., $N=2$ (an example would
be the $\rho^\pm$ system).
We can define as usual the eigenstates of positive and negative
charge by $\phi_\pm (x) = (\phi_1 (x) \pm i \, \phi_2 (x))/\sqrt{2}$.
Inserting these states into (\ref{gleich1}) we obtain the 
free Lagrangian
\begin{eqnarray} L^{\,0}_{\,\phi} (x)
 & = &
 (\partial_\mu \,\phi_{+} (x) ) (\partial^{\,\mu} \,\phi_{-} (x) )  - 
\, M^2 \; \phi_{+} (x) \, \phi_{-} (x) \; + \nonumber \\
 & + &
(\partial_\mu \,\phi_{-}^+ (x) ) (\partial^{\,\mu} \,\phi_{+}^+ (x) ) 
- \, M^{\ast \, 2} \;
\phi_{-}^+ (x)\, \phi_{+}^+ (x)\; .
\end{eqnarray}
Without loss of generality we have chosen an asymmetric field ordering,
which is very suitable for the quantisation of the system.
If we make the replacement $\phi_{-} (x) \rightarrow \phi_R (x)$ and $
\phi_{+} (x) \rightarrow \phi^+_L (x)$, we are back to the notation of
Ref.~\cite{Kle99}, but now for the charged bosonic
case. For the free Lagrangian we then obtain
\begin{eqnarray} L^{\,0}_{\,\phi} (x) & = &
 (\partial_\mu \,\phi^+_L (x) ) (\partial^{\,\mu} \,\phi_R (x) )  - \,
M^2 \;
\phi^+_L (x) \, \phi_R (x) \; + \nonumber \\
 & + &
(\partial_\mu \,\phi^+_R (x) ) (\partial^{\,\mu} \,\phi_L (x) )  - \,
M^{\ast \, 2} \;
\phi^+_R (x)\, \phi_L (x)\; .
\end{eqnarray}
The Lagrange equations of motion are now
\begin{equation}
(\,\partial^2 + M^2) \,\phi_\pm (x) \; = \; 0 \quad , \quad (\,
\partial^2 + M^{\ast \, 2}) \,\phi^+_\mp (x) \; = \; 0\; .
\end{equation}
The standard canonical momenta conjugate to the fields $\phi_\pm (x)$
and $\phi^+_\pm (x)$ are
\begin{equation} \Pi_\pm (x) = \frac{\delta \,L^{\,0}_{\,
\phi} (x)}{\delta \, (\partial_0 \,\phi_\pm (x))} = 
\partial_0 \,\phi_\mp (x) \, , \,\, \Pi_\pm^+ (x) = \frac{
\delta \,L^{\,0}_{\,\phi} (x)}{\delta \, (\partial_0 \,\phi_\pm^+ (x))}  = 
\partial_0 \,\phi_\mp^+ (x)\, .
\end{equation}
Canonical quantisation in configuration space yields
the {\em non-vanishing} \/equal-time commutation relations
\begin{eqnarray}
{ [ \, \phi_\pm (\vec{x},t) , \Pi_\pm (\vec{y},t) \, ] }
 & = &
 i\, \delta^{\, 3} (\vec{x} - \vec{y}\,)  \nonumber \\
 & & \nonumber \\
{ [ \, \phi^+_\pm (\vec{x},t) , \Pi^+_\pm (\vec{y},t) \, ] }
 & = & i\, \delta^{\, 3} (\vec{x} - \vec{y}\,)  \; .
\end{eqnarray}
The Laplace transformation for the charged field operators gives
\begin{eqnarray}
\phi_\pm (x) & = & \int \! \!
\frac{d^3 k}{(2\pi )^3 \; 2\, \omega \, (| \vec{k}\,|)}
\;\;
\Big[ \,
 a_\pm  (\vec{k}\,) \; e^{\displaystyle - \, i k x} \; + \;
 c_\mp^+  (\vec{k}\,) \; e^{\displaystyle i k x}
\,\Big] \nonumber \\
 & & \nonumber \\
\phi_\pm^+ (x) & = & \int \! \!
\frac{d^3 k}{(2\pi )^3 \; 2\,\omega^\ast (|\vec{k}\,|)}
\;\;
\Big[\,
  c_\mp  (\vec{k}\,) \; e^{\displaystyle - \, i k^\ast x} \; + \;
 a_\pm^+  (\vec{k}\,) \; e^{\displaystyle i k^\ast x}
\,\Big]\; ,
\end{eqnarray}
with the following {\em non-vanishing} \/commutation relations for
creation and annihilation operators in momentum space:
\begin{eqnarray}
{ [ \, a_\pm (\vec{k}\,) \; , \; c_\pm^+ (\vec{k}^{\,\prime}\,) \, ] }
 & = &
 (2\pi)^3 \, 2 \, \omega \,(|\vec{k}\,|)\; \delta^{\, 3} (\vec{k} - 
\vec{k}^{\,\prime}\,) \nonumber \\
 & & \nonumber \\
{ [ \, c_\pm (\vec{k}\,) \; , \; a_\pm^+ (\vec{k}^{\,\prime}\,) \, ] }
 & = &
 (2\pi)^3 \, 2 \, \omega^\ast(|\vec{k}\,|)\; \delta^{\, 3} (\vec{k} - 
\vec{k}^{\,\prime}\,)\; .
\end{eqnarray}
The Hamilton operator is
\begin{eqnarray}
H^{\,0}_{\phi} & = & \int\! d^{\,3}k \,\;
\frac{1}{2} \; \omega \,(|\vec{k}\,|)\; \Big( \,
c_{+}^+ (\vec{k}\,) \; a_{+} (\vec{k}\,) +
c_{-}^+ (\vec{k}\,) \; a_{-} (\vec{k}\,) \nonumber \\
 & & \qquad\qquad\qquad\quad
+ \, 
a_{+} (\vec{k}\,) \; c_{+}^+ (\vec{k}\,) +
a_{-} (\vec{k}\,) \; c_{-}^+ (\vec{k}\,)
\; \Big) \nonumber \\
 & + & \int\! d^{\,3}k \;
\frac{1}{2} \; \omega^\ast (|\vec{k}\,|)\; \Big( \,
a_{-}^+ (\vec{k}\,) \; c_{-} (\vec{k}\,) +
a_{+}^+ (\vec{k}\,) \; c_{+} (\vec{k}\,) \nonumber \\
 & & \qquad\qquad\qquad\quad
+ \, 
c_{-} (\vec{k}\,) \; a_{-}^+ (\vec{k}\,) +
c_{+} (\vec{k}\,) \; a_{+}^+ (\vec{k}\,)
\; \Big)\; .
\end{eqnarray}
Again keeping in mind that the words ``creation'' and ``annihilation''
are associated with the forward time direction, we can read off
the following identifications (starting with a positively charged particle):
\begin{eqnarray}
c_{+}^+ \quad & \leftrightarrow & \quad \mbox{creation operator of (charged)
bosonic particle}, \nonumber \\
c_{-}^+ \quad & \leftrightarrow & \quad \mbox{creation operator of (charged)
bosonic anti-particle}, \nonumber \\
a_{+}^+ \quad & \leftrightarrow & \quad \mbox{creation operator of (charged)
bosonic anti-hole}, \nonumber \\
a_{-}^+ \quad & \leftrightarrow & \quad \mbox{creation operator of (charged)
bosonic hole}, \nonumber \\
a_{+} \quad & \leftrightarrow & \quad \mbox{annihilation operator of (charged)
bosonic particle}, \nonumber \\
a_{-} \quad & \leftrightarrow & \quad \mbox{annihilation operator of (charged)
bosonic anti-particle}, \nonumber \\
c_{+} \quad & \leftrightarrow & \quad \mbox{annihilation operator of (charged)
bosonic anti-hole}, \nonumber \\
c_{-} \quad & \leftrightarrow & \quad \mbox{annihilation operator of (charged)
bosonic hole}\; .
\end{eqnarray}
Obviously, we can make the identifications $a_\pm = (a_1 \pm i \, a_2)/
\sqrt{2}\;$ and $\;c_\pm = (c_1 \pm i \, c_2)/\sqrt{2}$.
The Feynman propagator of a charged bosonic field reads
\begin{equation}
i \, \Delta_F (x-y)\; = \; <0|\,T\,(\,
\phi_{-}(x)\, \phi_{+}(y)\,)\,|0>  \; = \; i \int
\!\frac{d^{\,4}k}{(2\,\pi)^4}\;\; \frac{e^{-\,i\,k (x-y)}}{k^2 - M^2}
\; .
\end{equation}

\subsubsection{Introduction of resonant bosonic vector fields}  
In Ref.~\cite{Kle99} it was observed that in
the fermionic case a consistent redefinition of spinors is necessary.
Hence, in order to introduce resonant bosonic vector fields we must
similarly find a consistent formulation of polarisation vectors for
resonant vector mesons. The problematic point about the definition of
a polarisation operator for a resonant field becomes more obvious here 
than in the case of a spinor.
The reason is simple: one defines a momentum-dependent
polarisation vector of a vector particle (quasi-real mass $M=m-i\,
\varepsilon$) by boosting the constant rest-frame polarisation vector,
i.e., one defines the polarisation vectors by
\begin{eqnarray}
{\varepsilon}^{\,\mu \,\lambda_z} (\vec{p}\, ) & = &
({\varepsilon}^{\,0 \,\lambda_z} (\vec{p}\, ) \, , \,
\vec{\varepsilon}^{\;\lambda_z} (\vec{p}\, ) ) \, = \, \Lambda^\mu_{\;
\nu} (\vec{p}\,) \; {\varepsilon}^{\,\nu \,\lambda_z} (\vec{0}\, ) \nonumber \\
 & & \nonumber \\
 & = &
\left( \frac{\vec{p} \cdot \vec{\varepsilon}^{\;\lambda_z}}{M} \, , \,
\vec{\varepsilon}^{\;\lambda_z}
+ \frac{\vec{p} \cdot \vec{\varepsilon}^{\;\lambda_z}}{M \,(\omega(|
\vec{p}\,|) + M)} \,
\vec{p} \;
\right)\; .
\end{eqnarray}
Here, $\mu$ is a Lorentz index with $\mu =0,1,2,3$ and $\lambda_z=0,\pm 1$
is the polarisation index.
The properties of the three polarisation vectors of such a massive
vector boson are well known:
\begin{eqnarray}
 p_{\,\mu}\,\Big|_{p^0=\omega(|\vec{p}\,|)} \; {\varepsilon}^{\,\mu \,
\lambda_z} (\vec{p}\, ) \; & = & \; 0 \nonumber \\
 & & \nonumber \\
(-1)^{\lambda_z} \,{\varepsilon}^{\,-\,\lambda_z} (
\vec{p}\, ) \cdot
      {\varepsilon}^{\,\lambda_z'} (\vec{p}\, ) \; & = & \;
(-1)^{\lambda_z} \, {\varepsilon}^{\,\mu \,-\, \lambda_z} (\vec{p}\, ) 
\,
 {\varepsilon}_{\mu}^{\;\;\,\lambda_z'} (\vec{p}\, ) \; = \; - \,
{\delta}^{\,\lambda_z\lambda_z'}  \nonumber \\
 & & \nonumber \\
 \sum\limits_{\lambda_z}\; (-1)^{\lambda_z} \, {\varepsilon}^{\,\mu \,
\lambda_z} (\vec{p}\, ) \, {\varepsilon}^{\,\nu \,-\,\lambda_z} (
\vec{p}\, )
 \; & = & \; - \,
 g^{\,\mu \nu} + \frac{p^{\,\mu}p^{\,\nu}}{M^{\,2}} \; .
\end{eqnarray}
The question is now how to extend this description to a resonant
particle with complex mass $M=m-i\,\Gamma/2$. Or in other words: what
is the correct boost matrix for a resonant field? Is it the real boost
matrix based on velocities $\vec{\beta}$ or is it the complex
boost matrix based on momenta, energies, and complex masses? As the
definition of the orbital angular momentum is based on momenta ($
\vec{L}=\vec{r}\times\vec{p}$\,) and not on velocities, we prefer here to
choose the complex boost matrix based on momenta, energies, and complex
masses, i.e., we make the definition
\begin{eqnarray}
\Lambda^\mu_{\;\nu} (\vec{p}\,) \; = \;
\left(
\begin{array}{ccc}
\displaystyle \frac{\omega (| \vec{p}\,|)}{M} &
\displaystyle  &
\displaystyle \frac{{\vec{p}}^{\;T}}{M} \\
 & & \\
\displaystyle \frac{{\vec{p}}}{M} &
\displaystyle  &
\displaystyle 1_3 \; + \;
\frac{\vec{p}\;\;{\vec{p}}^{\;T}}{M \, (M + \omega (| \vec{p}\,|))}
\end{array}
\right)\; .
\end{eqnarray}
The consequence is, of course, that a boost of a resonance field
from a real space-time point will lead to complex space-time, which
is not so unnatural though, as a resonant particle will decay in the future,
i.e., it will disappear from space-time. Of course, it would be
interesting to discuss what is the meaning of the words
energy conservation (or, in general, the conservation of quantum
numbers), simultaneity, and causality for resonant fields, and to
find the answer to the question where in the complex space-time is
the rest frame of a moving resonant particle like the $\rho$ meson.
  
Returning to the introduction of bosonic field operators, we
can now consistently write down the Laplace transformation of bosonic
resonant vector field operators like the $\rho$ meson 
($\chi_{\,t_z}$ represent the isospinors):
\begin{eqnarray}
\phi^{\,\mu} (x) \; = \; \sum\limits_{t_z} \;\sum\limits_{\lambda_z} & &  
\int \! \!
\frac{d^3 k}{(2\pi )^3 \; 2\, \omega \, (| \vec{k}\,|)} \nonumber \\
 & &
\Big[ \,  
 {\varepsilon}^{\,\mu \,\lambda_z} (\vec{k}\, ) \; \chi_{\,t_z} \; a \,
(\vec{k} , \lambda_z,t_z) \; e^{\displaystyle - \, i k x} \nonumber \\
 & & + \, 
(-1)^{\lambda_z} \, {\varepsilon}^{\,\mu \,- \lambda_z} (\vec{k}\, ) \;
\chi^+_{\,t_z}  \; c^+ \, (\vec{k} , \lambda_z,t_z) \; e^{\displaystyle
i k x}
\,\Big] \nonumber \\
 & & \nonumber \\
\phi^{\,+\,\mu} (x) \; = \; \sum\limits_{t_z} \;\sum\limits_{\lambda_z} & &  
\int \! \!
\frac{d^3 k}{(2\pi )^3 \; 2\,\omega^\ast (|\vec{k}\,|)} \nonumber \\
 & &
\Big[\,
(-1)^{\lambda_z}  ({\varepsilon}^{\,\mu \,- \lambda_z} (\vec{k}\, ))^
\ast \; \chi_{\, t_z} \;  c \, (\vec{k} , \lambda_z,t_z) \; e^{
\displaystyle - \, i k^\ast x}\nonumber \\
 & & \quad \qquad \,+
 ({\varepsilon}^{\,\mu \,\lambda_z} (\vec{k}\, ))^\ast \; \chi^+_{
\,t_z} \;  a^+ \, (\vec{k} , \lambda_z,t_z) \; e^{\displaystyle i k^
\ast x}\,\Big]\; . 
\end{eqnarray}
Using these field operators, it is now straightforward to calculate
the corresponding Feynman propagator ($\phi^{\,\mu} \,(x) =  \sum
\limits_{t_z} \;\sum\limits_{\lambda_z} \; \phi_{\,t_z}^{\,\mu\,
\lambda_z}\,(x)$ ):
\begin{eqnarray} \lefteqn{i \, \Delta^{\,\mu\nu}_F (x-y) \;\;  \delta^{\,
\lambda_z\lambda_z'} \;\;  \delta_{\,t_z t_z'} \; = \;  <0|\,T\,(\,
\phi_{\,t_z}^{\,\mu\,\lambda_z}\,(x)\, \phi_{\,t_z'}^{\,\nu\,
\lambda_z'}\, (y)\,)\,|0> \; =} \nonumber \\
 & & \nonumber \\
 & = & i \int\!\frac{d^{\,4}k}{(2\,\pi)^4}\;\; \frac{e^{-
\,i\,k (x-y)}}{k^2 - M^2} \;\; \left(  - \,
 g^{\,\mu \nu} + \frac{k^{\,\mu}k^{\,\nu}}{M^{\,2}} \right) \;\; 
\delta^{\,\lambda_z\lambda_z'} \;\;  \delta_{\,t_z t_z'}\; .
\end{eqnarray}  
The quantisation of resonant vector fields is analogous to the
quantisation of resonant scalar fields.
\clearpage

\subsection{Interactions}
Following the discussion of the free QFTs, it is now interesting to study
the consistent implementation of interactions.
Let us start with a quasi-real (uncharged) bosonic QFT
($N=1$), for which we consider the interaction
($\phi^{3}$-theory) Lagrangian
\begin{eqnarray}
L_{\,\phi} (x) \; & = &
L^{\,0}_{\,\phi} (x) \; + \; L^{\,int}_\phi (x) \nonumber
\\
 & & \nonumber \\
L^{\,0}_{\,\phi} (x) \; & = &
\frac{1}{2} \left( (\partial_\mu \,\phi (x) ) (\partial^{\,\mu} \,\phi
(x) )  - \, M^2 \,
\phi (x) \, \phi (x) \right) \; + \nonumber \\
 & + &
\frac{1}{2} \left( (\partial_\mu \,\phi^+ (x) ) (\partial^{\,\mu} \,
\phi^+ (x) )  - \, M^{\ast \, 2} \,
\phi^+ (x)\, \phi^+ (x) \right)
\nonumber \\
 & & \nonumber \\
L^{\,int}_\phi (x) \; & = & \;
- \, \frac{1}{3!} \,\; g_3 \;\, \phi^3 (x) \, \quad \; - \;
 \frac{1}{2!} \; g_{2,1} \; \phi^2(x) \;\, \phi^+ (x) \nonumber \\
 & & \; - \;
 \frac{1}{3!} \; g^\ast_3 \; (\phi^+ (x))^3 \;  - \;
 \frac{1}{2!} \; g^\ast_{2,1} \; \phi\, (x) \; (\phi^+ (x))^2  \; ,
\label{Lagr1}
\end{eqnarray}
where $g_3$ and $g_{\,2,1}$ are, for the moment, arbitrary
{\em complex} coupling constants.  
The interaction terms containing both $\phi$ and $\phi^+$ fields
need special consideration, as they yield direct interactions between
(anti-)particles and (anti-)holes. The consequence of these terms is
that e.g.\ a particle can lower its energy by creating an infinite
number of holes, whereas a hole can increase its energy by creating an
infinite number of particles. This situation is commonly called
``radiation catastrophe'' and has been cured in the fermionic case by the
introduction of a Dirac sea, whereas in the bosonic case the
Bose-Einstein condensation of particles to infinite negative energies
has been avoided by forbidding the existence of the negative-energy
states of the Klein-Gordon equation. This asymmetric and arbitrary
treatment of the problem of fermionic and bosonic negative-energy
states in QFT is quite unsatisfactory, as was pointed out e.g.\ in
Chapter 1 of \cite{Wei95}.

In order to avoid the ``radiation catastrophe'' in a way which is symmetric
for fermions and bosons, the following rule has to be made
up\footnote{With respect to Ref.~\cite{Kle99},
the rule is equivalent to demanding the diagonal elements of the vertex
matrix to be zero.}: \\[4mm]
\framebox{{\em Direct interactions between (anti-)particles and
(anti-)holes aren't allowed!}}\\[4mm]
As a consequence for the Lagrangian under consideration,
formula (\ref{Lagr1}), we have to set $g_{2,1} = g^\ast_{2,1} = 0$.
\clearpage
  
\section{The \bm{\omega (782)\rightarrow 3\pi} partial decay width}
\label{UERMresult}
\subsection{The standard lowest-order chiral Lagrangian of the
\bm{\omega\rho\pi} system}
The $\omega\rho\pi$ system can be described in a simple
meson picture by the phenomenological ``effective'' Lagrangian
\begin{equation} L (x) \; = \;
L^{\,0}_{\,\omega} (x) +
L^{\,0}_{\,\rho} (x) +
L^{\,0}_{\,\pi} (x) +
L^{\,int}_{\,\omega\,\leftrightarrow \,\rho\,\pi} (x) +
L^{\,int}_{\,\rho\,\leftrightarrow \,\pi\pi} (x) + \ldots
\; .
\label{Lag1}
\end{equation}
For the determination of the respective Lagrangians, we will follow 
Ref.~\cite{Bra95}, based on the ``hidden-symmetry approach'' of
Ref.~\cite{Ban85}.
First we define the SU(3)-octet matrix for the pseudoscalar mesons by: 
\begin{eqnarray}
P & \simeq & \left(
\begin{array}{ccccc}
\displaystyle \frac{\pi^0}{\sqrt{2}} + \frac{\eta_8}{\sqrt{6}} & & 
\pi^+ & & K^+ \\
\pi^- & & \displaystyle -\, \frac{\pi^0}{\sqrt{2}} + \frac{\eta_8}{
\sqrt{6}} & & K^0 \\
K^- & & \bar{K}^0 & & \displaystyle -\, \frac{2\,\eta_8}{\sqrt{6}}
\end{array}
\right) \; .  
\end{eqnarray}
Here we made the reasonable approximation $\eta_3 \simeq \pi^0$, i.e.,
we assume that $\eta_8$ does not contain any $\pi^0$-component.
The corresponding matrix for the ``ideally mixed'' vector-meson-octet --- 
``ideal mixing'' is suggested to good accuracy by nature (see e.g. the
{\it Quark Model} section of Ref.~\cite{PP}) ---
can be introduced by:
\begin{eqnarray}
V  & \simeq & 
\left(
\begin{array}{ccccc}
\displaystyle \frac{\rho^0}{\sqrt{2}} + \frac{\omega}{\sqrt{2}} & &
\rho^+ & & K^{\,\ast \, +} \\
\rho^- & & \displaystyle -\, \frac{\rho^0}{\sqrt{2}} + \frac{
\omega}{\sqrt{2}} & & K^{\,\ast \, 0} \\
K^{\,\ast \, -} & & \bar{K}^{\,\ast\, 0} & & \displaystyle \phi
\end{array}
\right)\; .
\end{eqnarray}
The lowest-order interactions between VPP and
PVV systems in the ``hidden-symmetry'' Lagrangian is described by the
terms ($\varepsilon^{\, 0123}=1$)
\begin{eqnarray}
L^{\,int}_{\,VPP} (x) \; & = & \;
i \, g \; \mbox{Tr}\Big[ \,V_\mu \, P \, (\partial^{\,\mu} P) - V_\mu 
\, (\partial^{\,\mu} P) \, P \,\Big] \nonumber \\
L^{\,int}_{\,VVP} (x) \; & = & \;
 \frac{G_\pi}{\sqrt{2}} \; \varepsilon^{\,\mu\nu\alpha\beta}\; 
\mbox{Tr}\Big[ \,(\partial_{\,\mu}\, V_\nu) \, (\partial_{\,\alpha} V_
\beta) P \,\Big]
\; .\end{eqnarray}
For the interaction in the $\omega\rho\pi$ system, these Lagrangians
yield for ``ideal mixing'' of the neutral vector mesons involved
\begin{eqnarray}
L^{\,int}_{\,\rho\pi\pi} (x) \; & = & \;
-\, \sqrt{2} \, g \; \vec{\rho}_\mu (x) \, \cdot \Big[ \,\vec{\pi} 
\,(x) \times (\partial^{\,\mu} \vec{\pi} \,(x)) \,\Big] \nonumber \\
L^{\,int}_{\,\omega \rho\pi} (x) \; & \simeq & \;
G_\pi \; \varepsilon^{\,\mu\nu\alpha\beta}\; (
\partial_{\,\mu}\, \omega_{\,\nu} (x)) \; (\partial_{\,\alpha} \vec{
\rho}_\beta \, (x)) \cdot \vec{\pi} \, (x)
\; ,\end{eqnarray}
Henceforth we will use for couplings the notation of Ref.~\cite{Bra95},
i.e., $g = g_{\rho\pi\pi} /\sqrt{2}$, $f = \sqrt{2}\, F_\pi \simeq 132$ MeV
with $F_\pi \simeq 93$ MeV and
$G_\pi = \sqrt{2}\, g \, g_{\omega\pi^0\gamma}$. 
The order of accuracy of the relation between the couplings and the
vector-meson masses of the ``hidden-symmetry'' Lagrangian,
i.e., $m^2_\rho \simeq m^2_\omega \simeq m^2_V = 2\, g^2 f^2
\simeq 0.60$ GeV${}^2$, suggests to what accuracy our later results
have to be interpreted.

Lagrangian (\ref{Lag1}) is quasi-hermitian and consists of
quasi-real meson fields only.
If one tries to take into account the observed finite
decay width $\Gamma_\rho$ of the $\rho$ meson into two pions, one will
have to include higher-loop terms up to infinite order in
perturbation theory to simulate the resonant $\rho$ intermediate
state.

\subsection{Determination of \bm{\Gamma(\omega\rightarrow\rho\pi
\rightarrow\pi\pi\pi)} from an effective Lagrangian based on the
well-defined QFT}
In order to perform a tree-level calculation without
loops\footnote{This is in the spirit of the loop shrinking that appears in
\cite{Sca00} in the treatment of the L$\sigma$M Lagrangian and the
extraction of finite information out of its lowest-order loop
diagrams.}, which is containing the essential information of the
calculation within a quasi-real theory to all orders of perturbation
theory, one can replace the Lagrangian of quasi-real fields by an
effective Lagrangian containing resonant fields, which then have to be
treated consistently within the well-defined QFT.

In practical calculations, one may obtain such a physically
realistic $\rho$ resonance in a coupled-channel approach, by also including
several virtual two-meson channels, as for instance
is done in Ref.~\cite{Bev83}. This $\rho$ resonance will then serve
as an input in the effective-Lagrangian approach to calculate the
leading-order $\omega$ decay width.

The $\rho$ meson is described within such an effective Lagrangian by a
resonant degree of freedom with a complex constant\footnote{This
assumption of course neglects the momentum dependence of the pole
paramaters $m_\rho$ and $\Gamma_\rho$.} ``mass'' $M_\rho = m_\rho - i
\Gamma_{\rho}/2$, whereas the $\omega$ and $\pi$ mesons, due to
their comparatively small decay widths, can be treated as quasi-real
fields.
The effective Lagrangian of the $\omega\rho\pi$ system in the
well-defined QFT will then be
\begin{eqnarray}
L^{\,\prime} (x) & = &
\Bigg( - \, \frac{1}{4} \, \omega_{\mu\nu} \,(x) \, \omega^{\,\mu\nu} \,(x) + 
\, \frac{1}{2} \; (m_{\,\omega}-i\,\varepsilon)^2 \;\;
\omega_{\mu} (x) \, \omega^{\,\mu} (x) \nonumber \\
 & &
- \, \frac{1}{4} \, \vec{\rho}_{\,\mu\nu} \,(x) \,\cdot \, \vec{\rho}^{
\,\,\mu\nu} \,(x) + \, \frac{1}{2} \; M^2_{\,\rho} \;\;
\vec{\rho}_{\mu} (x) \,\cdot\, \vec{\rho}^{\,\mu} (x)  \nonumber \\
 & &
+ \, \frac{1}{2} \left( (\partial_\mu \,\vec{\pi}\,(x) ) \cdot (
\partial^{\,\mu} \, \vec{\pi}\,(x) )  - \, (m_\pi-i\,\varepsilon)^2 \;
\;
\vec{\pi}\, (x) \,\cdot\, \vec{\pi}\, (x) \right)  \nonumber \\
 & & + \,
G^{\,\prime}_\pi \; \varepsilon^{\,\mu\nu\alpha\beta}\; (\partial_{\,
\mu}\, \omega_{\nu} (x)) \; (\partial_{\,\alpha} \vec{\rho}_\beta \,
(x)) \cdot \vec{\pi} \, (x) \nonumber \\
 & & -\, \sqrt{2} \, g^{\,\prime} \; \vec{\rho}_\mu (x) \, \cdot \Big[ \,
\vec{\pi} \,(x) \times (\partial^{\,\mu} \vec{\pi} \,(x)) \,\Big] 
 + \ldots \Bigg) + \mbox{h.c.}
\; .\end{eqnarray}
Some important questions are now:
How will the parameters $G^{\,\prime}_\pi$ and $g^{\,\prime}$ of the
effective Lagrangian change when compared to the original Lagrangian? Will
they develop a complex phase? What are the compensating interaction
terms in the effective Lagrangian which guarantee the quasi-reality
of the effective action in the (anti-)particle  and the
(anti-)hole sector.
How will the phase-space integral change, if there are resonant fields
in the final state?
Some of these questions we try to answer in the following discussion. For
simplicity we choose $G^{\,\prime}_\pi \simeq G_\pi$ and $g^{\,
\prime} \simeq g$. If the thus calculated partial $\omega$ width
into three pions turns out to be within the range of the experimental
measurements, then this assumption should be reasonable, otherwise it will
have to be reconsidered.

Using the introduced effective Lagrangian $L^{\,\prime} (x)$, the
partial decay width $\Gamma(\omega\rightarrow\rho\pi\rightarrow \pi\pi
\pi)$ is calculated to lowest, though leading, order by
(see e.g.\ p.\ 80 in Ref.~\cite{Nac91})($s = p^2 = m_\omega^2$)
\begin{eqnarray}
\lefteqn{\Gamma_{\omega\rightarrow\rho\pi\rightarrow \pi\pi\pi} (
\vec{p}\,) \; \simeq \; \frac{1}{2\, \sqrt{s}} \;\frac{1}{3!} \;  
\frac{1}{3} \, \sum\limits_{\lambda_z} \; \sum\limits_{t_{z1}}\; \sum
\limits_{t_{z2}}\; \sum\limits_{t_{z3}}} \nonumber \\
 & & \nonumber \\
 & & \frac{1}{(2\pi)^5}
\int \frac{d^3k_1}{2\,\omega_{\pi} (|\vec{k}_1|)} \;
\frac{d^3k_2}{2\,\omega_{\pi} (|\vec{k}_2|)} \;
\frac{d^3k_3}{2\,\omega_{\pi} (|\vec{k}_3|)} \;\;
 \delta^4 (k_1 + k_2 + k_3 - p) \nonumber \\
 & & \nonumber \\
 & &
|<\pi(\vec{k}_1, t_{z1})\,\pi(\vec{k}_2, t_{z2})\,\pi(\vec{k}_3,
t_{z3})|\,T \Big[ i^2  :L^{\,\prime\,int}_{\,\rho\pi\pi} (0):
\; :S^{\,\prime\,int}_{\,\omega\rho\pi}: \Big]\, |\,\omega(
\vec{p},\lambda_z)> |^2
\; . \nonumber \\
\end{eqnarray}
To be more explicit, the in- and out-state vectors in terms of
creation and annihilation operators are given by
\begin{eqnarray}
<\pi(\vec{k}_1, t_{z1})\,\pi(\vec{k}_2, t_{z2})\,\pi(\vec{k}_3,
t_{z3})| & = & <0| \, a_{\,\pi}(\vec{k}_3, t_{z3})\;a_{\,\pi}(
\vec{k}_2, t_{z2})\;a_{\,\pi}(\vec{k}_1, t_{z1}) \nonumber \\
 & & \nonumber \\
 |\,\omega(\vec{p},\lambda_z)> & = & c^+_{\,\omega}(
\vec{p},\lambda_z) \,|0>
\; .\end{eqnarray}
The T matrix of the process can be simplified by Wick's
theorem\footnote{Within the developed formalism for resonant fields, the proof
of the validity of Wick's theorem is straightforward, but for the
(anti-)hole sector of the theory one has to replace the time-ordered
particle propagators by the anti-time-ordered hole propagators.}:
\begin{eqnarray} \lefteqn{i\, T_{fi}\; =\;
<\pi(\vec{k}_1, t_{z1})\,\pi(\vec{k}_2, t_{z2})\,\pi(\vec{k}_3,
t_{z3})|T \Big[ i^2\! :L^{\,\prime\,int}_{\,\rho\pi\pi} (0):
 : S^{\,\prime\,int}_{\,\omega\rho\pi}: \Big] |\,\omega(
\vec{p},\lambda_z)>} \nonumber \\ [.3cm] & = &
i^2 \int d^4z \,<\pi(\vec{k}_1, t_{z1})\,\pi(\vec{k}_2, t_{z2})\,\pi(
\vec{k}_3, t_{z3})|\nonumber \\ [.3cm] \lefteqn{T \Big[
 :\sqrt{2} \, g^{\,\prime} \vec{\rho}_{\bar{\mu}} (x) \cdot \Big[
(\partial^{\,\bar{\mu}} \vec{\pi} \,(x) \times \vec{\pi} \,(x))
\Big]\Big|_{x=0} : :G^{\,\prime}_\pi
\varepsilon^{\,\mu\nu\alpha\beta}\, (\partial_{\,\mu}\, \omega_{\nu}
(z)) (\partial_{\,\alpha} \vec{\rho}_\beta \, (z)) \cdot \vec{\pi}
\, (z): \Big] \, |\,\omega(\vec{p},\lambda_z)>} \nonumber \\ [.3cm]
 & = & i^2 \; \int d^{\,4}z \;\; e^{i\,x (k_1+k_2) + i \, z (k_3-\,p) }
\; i \int\!\frac{d^{\,4}k}{(2\,\pi)^4}\;\;
\frac{e^{-i\,k (x-z)}}{k^2
- M_\rho^2} \;\; \left(  - \,
 g_{\,\bar{\mu} \beta} + \frac{k_{\,\bar{\mu}}k_{\,\beta}}{M_\rho^{
\,2}} \right) \nonumber \\ [.3cm]
 & & \;\;\;\;\;\;\;\;\;\;\;\;\;\;\;\;
 \sqrt{2} \, g^{\,\prime} \; G^{\,\prime}_\pi \;
\varepsilon^{\,\mu\nu\alpha\beta}\; \varepsilon_{\,t_{z1}\,t_{z2}\,t_{z3}}\;
 i\, (k^{\,\bar{\mu}}_1 - k^{\,\bar{\mu}}_2) (-\,i)\, p_{\,\mu}\,
\varepsilon^{\,\lambda_z}_{
\,\nu} (\vec{p}\,)_\omega \; (i\, k_{\,\alpha}) \Big|_{x=0}\; +
 \nonumber \\ [.3cm]
 & + & \quad \mbox{terms obtained by cyclic permutation of indices 1, 2 and 3}
 \nonumber \\ [.3cm] & = &
\int d^{\,4}k \;\; \int\!\frac{d^{\,4}z}{(2\,\pi)^4}\;\; e^{i \, z
(k + k_3-\,p) } \;\; \frac{1}{k^2 - M_\rho^2}\;
\left(  - \, k_{1\, \beta} + k_{2\,
\beta} + \frac{(k_1-k_2)\cdot k \;  k_{\,\beta}}{M_\rho^{\,2}} \right)
\nonumber \\ [.3cm] & & \quad\qquad\qquad\qquad\qquad\qquad
\sqrt{2} \, g^{\,\prime} \; G^{\,\prime}_\pi \; \varepsilon_{\,t_{z1}
\,t_{z2}\,t_{z3}} \;\varepsilon^{\,\mu\nu\alpha\beta}\;  p_{\,\mu}\,
\varepsilon^{\,\lambda_z}_{\,\nu} (\vec{p}\,)_\omega \;  k_{\,\alpha}
\nonumber \\ [.3cm]
 & + & \quad \mbox{terms obtained by cyclic permutation of indices 1, 2 and 3}
 \nonumber \\ [.3cm] &  = &
\int d^{\,4}k \; \delta^4(k + k_3-\,p) \; \frac{1}{k^2 - M_
\rho^2} \; \left(  - \, k_{1\, \beta} + \, k_{2\, \beta}\right)\;
\sqrt{2} \, g^{\,\prime} G^{\,\prime}_\pi \, \varepsilon_{\,t_{z1}
\,t_{z2}\,t_{z3}} \,\varepsilon^{\,\mu\nu\alpha\beta}\,  p_{\,\mu}\, 
\varepsilon^{\,\lambda_z}_{\,\nu} (\vec{p}\,)_\omega \;  k_{\,\alpha} 
\nonumber \\ [.3cm] & + &
\quad \mbox{terms obtained by cyclic permutation of indices 1, 2 and 3}
\nonumber \\ [.3cm] & = &
\sqrt{2} \, g^{\,\prime} \; G^{\,\prime}_\pi \; \varepsilon_{
\,t_{z1}\,t_{z2}\,t_{z3}} \; \varepsilon^{\,\mu\nu\alpha\beta}\;
p_{\,\mu}\, \varepsilon^{\,\lambda_z}_{\,\nu} (\vec{p}\,)_\omega \nonumber \\
[.3cm] & & \;\;\;\;\;\;\;\;\;\;\;\;\;\;\left( \;
\frac{k_{3\,\alpha} \; (k_{1\, \beta} - k_{2\, \beta})}{(p-\,k_3)^2 - M_\rho^2}
+ \frac{k_{1\,\alpha} \; (k_{2\, \beta} - k_{3\, \beta})}{(p-\,k_1)^2
- M_\rho^2}+ \frac{k_{2\,\alpha} \; (k_{3\, \beta} -
k_{1\, \beta})}{(p-\,k_2)^2 - M_\rho^2} \; \right)
\; .\end{eqnarray}
Before proceeding it is useful to make a short remark: so far all
partial derivatives acting on the resonant $\rho$ fields were acting
directly on the exponential function of the Laplace transform of the $
\rho$ propagator. As a consequence, the partial derivatives were
replaced by $i$ times plus or minus some {real} momentum. Therefore, we
had no difficulties afterwards to replace the configuration integrals
over these exponentials by four-momentum-conserving 
$\delta$-distributions and to perform the final four-momentum
integration in the variable $k$. {\em However, if we had some resonant
fields in the initial or final states, the derivatives would act
on the plane-wave exponentials of these fields, which would lead to a
replacement of the derivatives by complex four-momenta. Even worse:
the plane-wave exponentials would enter the configuration integrals, which
would give four-momentum-conserving $\delta$-distributions. As they
would now contain complex four-momenta, the resulting integrals would no longer
be simple $\delta$-distributions, i.e., they could not be easily integrated
away.}
The conclusion is that, for {\em intermediate} \/resonance fields we have no
difficulties with the four-momentum-conserving
$\delta$-distributions, whereas for resonance fields in the {\em initial} \/or
{\em final} \/state we run into problems. Even worse, the overall
four-momentum conservation in the phase-space integral would have to be
reformulated for resonant fields in the final state.
The next step is to perform the spin-isospin averaging of $|i
\,T_{fi}|^2$. We obtain:
\begin{eqnarray} \lefteqn{\overline{|i\, T_{fi}|^2} \; = \; \frac{1}{3!} \;  
\frac{1}{3} \, \sum\limits_{\lambda_z} \; \sum\limits_{t_{z1}}\; \sum
\limits_{t_{z2}}\; \sum\limits_{t_{z3}}\; |i\,T_{fi}|^2\; =\;
\frac{1}{3} \, \sum\limits_{\lambda_z} \;\Bigg| \,
\sqrt{2} \, g^{\,\prime} \; G^{\,\prime}_\pi \;  
\varepsilon^{\,\mu\nu\alpha\beta}\; p_{\,\mu}\, \varepsilon^{\,
\lambda_z}_{\,\nu} (\vec{p}\,)_\omega} \nonumber \\
 & & \quad
\left( \;
\frac{k_{3\,\alpha} \; (k_{1\, \beta} - k_{2\, \beta})}{(p-\,k_3)^2 - M_\rho^2}
+ \frac{k_{1\,\alpha} \; (k_{2\, \beta} - k_{3\,
\beta})}{(p-\,k_1)^2 - M_\rho^2} + \frac{k_{2\,\alpha} \; (k_{3\, \beta}
- k_{1\, \beta})}{(p-\,k_2)^2 - M_\rho^2} \; \right) \Bigg|^2
\; .\;\;\;\;\;\;\;\;\;\;\end{eqnarray}
At this point we again have to make a remark. Only because the decaying
$\omega$ meson is treated as a real particle, i.e., $\Gamma_\omega = 
\varepsilon$, we can use the following identity for its polarisation
vector:
\begin{equation} \sum\limits_{\lambda_z} \;\;\varepsilon^{\,
\lambda_z}_{\,\nu} (\vec{p}\,)_\omega \;\;(\varepsilon^{\,\lambda_z}_{
\,\bar{\nu}} (\vec{p}\,)_\omega)^\ast \; = \; -\, g_{\,\nu\,\bar{\nu}}
+ \frac{p_{\nu}\,p_{\bar{\nu}}}{m^2_{\,\omega}}
\; .\end{equation}
As already described above, in general the relation $(\varepsilon^{\,
\lambda_z}_{\,\mu} (\vec{p}\,))^\ast = (-1)^{\,
\lambda_z} \;\varepsilon^{\,- \lambda_z}_{\,\mu} (\vec{p}\,)$ is no
longer valid for objects with complex mass and has to be modified.  
Using these results we continue the spin-isospin averaging.
In the end, using the symmetry of the phase-space integral under arbitrary
permutations of the three {\em mass degenerate} pions in the final state we
obtain the following expression for the partial width:
\begin{eqnarray}
\lefteqn{\Gamma_{\omega\rightarrow\rho\pi\rightarrow \pi\pi\pi} (\vec{p}\,)
\;\simeq\;\frac{|
 \sqrt{2} \, g^{\,\prime} \; G^{\,\prime}_\pi |^2}{2\, \sqrt{s} \; (2\pi)^5}
\int \frac{d^3k_1}{2\,\omega_{\pi} (|\vec{k}_1|)} \;
\frac{d^3k_2}{2\,\omega_{\pi} (|\vec{k}_2|)} \;
\frac{d^3k_3}{2\,\omega_{\pi} (|\vec{k}_3|)} \;\; \delta^4
(k_1 + k_2 + k_3 - p)} \nonumber \\ [.3cm]
& & \Bigg\{ \;\left|
\begin{array}{ccccc}
p^2 & & k_3 \cdot p & & (k_1 - k_2) \cdot p \\
p \cdot k_3 & & k_3^2 & & (k_1-k_2) \cdot k_3 \\
p \cdot (k_1-k_2) & & k_3 \cdot (k_1 - k_2) & & (k_1 - k_2)^2 
\end{array}
\right| \;\; \frac{1}{|(p-\,k_3)^2 - M_\rho^2 \, |^2}\;\;\;\;\;\;\;\;\;\;
\nonumber \\ [.3cm] & & + \, 
\left|
\begin{array}{ccccc}
p^2 & & k_3 \cdot p & & (k_1 - k_2) \cdot p \\
p \cdot k_1 & & k_3 \cdot k_1 & & (k_1 - k_2) \cdot k_1 \\
p \cdot (k_2 - k_3) & & k_3 \cdot (k_2 - k_3) & & 
(k_1 - k_2) \cdot (k_2 - k_3)
\end{array}
\right|  
\nonumber \\
 & & \nonumber \\
 & & \qquad\qquad\qquad\quad
 2\;\mbox{Re} \, \left[
 \frac{1}{((p-\,k_3)^2 - M_\rho^2)\;((p-\,k_1)^2 - M_\rho^{\ast\,2})}
\right] \; \Bigg\} 
\; .\end{eqnarray}
\clearpage

Using four momentum contraints provided by the phase-space integral both
determinats can be brought to the same form yielding:
\begin{eqnarray}
\lefteqn{\Gamma_{\omega\rightarrow\rho\pi\rightarrow \pi\pi\pi} (\vec{p}\,)
\;\simeq\;\frac{|
 \sqrt{2} \, g^{\,\prime} \; G^{\,\prime}_\pi |^2}{2\, \sqrt{s} \; (2\pi)^5}
\int \frac{d^3k_1}{2\,\omega_{\pi} (|\vec{k}_1|)} \;
\frac{d^3k_2}{2\,\omega_{\pi} (|\vec{k}_2|)} \;
\frac{d^3k_3}{2\,\omega_{\pi} (|\vec{k}_3|)} \;\;
\delta^4 (k_1 + k_2 + k_3 - p)} \nonumber \\ [.3cm] & & 
4 \, \left|
\begin{array}{ccccc}
p^2 & & k_3 \cdot p & & p \cdot k_1 \\
p \cdot k_3 & & m_\pi^2 & & k_1 \cdot k_3 \\
p \cdot k_1 & & k_1 \cdot k_3 & & m_\pi^2 
\end{array}
\right|\nonumber \\ [.3cm] & & \; \Bigg\{ \; 
 \frac{1}{| (p-\,k_3)^2 - M_\rho^2\,|^2} +
 2\;\mbox{Re} \, \left[
 \frac{1}{((p-\,k_3)^2 - M_\rho^2)\;((p-\,k_1)^2 - M_\rho^{\ast\,2})}
\right] \; \Bigg\} 
\; .\;\;\;\;\;\;\;\;\;\;\nonumber \\
\end{eqnarray}

In order to evaluate the phase-space integral, it is useful to introduce
the Lorentz invariants $s = p^2 = m^2_\omega$ and
\begin{equation} s_1 = (k_1+k_2)^2
= (p-\,k_3)^3 \quad , \quad s_2 = (k_2 + k_3)^2 =  (p-\,k_1)^2
\; .\end{equation}
The following Lorentz-invariant scalar products of four-momenta are relevant
to our calculation:
\begin{eqnarray}
 & 2 & p \cdot k_1 = m_\pi^2 + s - \, s_2 \;\; , \;\;
2\, p \cdot k_3 = m^2_\pi + s - s_1 \;\; , \nonumber \\
 & & \nonumber \\
 & 2 &  k_1 \cdot k_3 = m_\pi^2 + s - \, s_1 - s_2
\; .\end{eqnarray}
Application of these scalar products yields the following expression for
the decay width:
\begin{eqnarray}
\lefteqn{\Gamma_{\omega\rightarrow\rho\pi\rightarrow \pi\pi\pi} (\vec{p}\,)
\;\simeq\;\frac{|
\sqrt{2} \, g^{\,\prime} \; G^{\,\prime}_\pi |^2}{2\, \sqrt{s} \; (2\pi)^5} \;
\; \frac{\pi^2}{4\, s} \int^{\,(\sqrt{s} - \, m_\pi)^2}_{\,(2\, m_
\pi)^2} \!\! ds_1\int^{\, s^+_2}_{\, s^-_2} ds_2} \\ [.3cm]
& & \Big[ \,
- s^2_1 \, s_2 + s_1 \, s_2 (s -s_2 + 3\, m^2_\pi ) - m^2_\pi (s-\,
m^2_\pi)^2
\; \Big]\;\;\;\;\;\;\;\;\;\;\;\nonumber \\ [.3cm] & &
\Bigg\{ \; \frac{1}{| s_1 - M_\rho^2 |^2} + 
  2\;\mbox{Re} \left[
 \frac{1}{(s_1 - M_\rho^2)\;(s_2 - M_\rho^{\ast\,2})}
\right]
 \; \Bigg\}
\nonumber 
\end{eqnarray}
or
\begin{eqnarray}
\lefteqn{\Gamma_{\omega\rightarrow\rho\pi\rightarrow \pi\pi\pi} (\vec{p}\,)
\;\simeq\;\frac{|
 \sqrt{2} \, g^{\,\prime} \; G^{\,\prime}_\pi |^2}{2\, \sqrt{s} \; (2\pi)^5} \;
\; \frac{\pi^2}{4\, s} \int^{\,(\sqrt{s} - \, m_\pi)^2}_{\,(2\, m_
\pi)^2} \!\! ds_1 \; 
\frac{1}{| s_1 - M_\rho^2 |^2}
\int^{\, s^+_2}_{\, s^-_2} ds_2} \\ [.3cm]
& & \Big[ \,
- s^2_1 \, s_2 + s_1 \, s_2 (s -s_2 + 3\, m^2_\pi ) - m^2_\pi (s-\,
m^2_\pi)^2 \; \Big]\;\;\;\;\;\;\;\;\;\;\;\;\nonumber \\ [.3cm] & &
\Bigg\{ \; 1 + \, 2\, 
 \frac{\left(s_2 - \mbox{Re} \left[ M_\rho^2 \right]\right)\left(s_1 - 
\mbox{Re} \left[ M_\rho^2 \right]\right) + \mbox{Im}^2\! 
\left[ M_\rho^2 \right]}{\left(s_2 - \mbox{Re} 
\left[ M_\rho^2 \right]\right)\left(s_2 - \mbox{Re} 
\left[ M_\rho^2 \right]\right) + \mbox{Im}^2\! \left[ M_\rho^2 \right]}
 \; \Bigg\} \; .
\nonumber
\end{eqnarray}
with ($\lambda (x,y,z) = x^2 + y^2 + z^2 - \, 2\, x\,y - \, 2 \, y\,z
- \, 2 \, z\,x $)
\begin{eqnarray}
 s^\pm_2 & = & \frac{1}{2\,s_1} \;
 \left[ \; s_1 \; (s-s_1+3\, m^2_\pi ) \; \pm \;
 \sqrt{\lambda (s,m^2_\pi ,s_1) \, \lambda (s_1 , m^2_\pi , m^2_\pi )} 
\; \right] \;\; , \nonumber \\
 & & \nonumber \\
 M_\rho & = & m_\rho - i \; \frac{\Gamma_\rho}{2} \;\; , \;\; \mbox{Re} 
\left[ M_\rho^2 \right] = m^2_\rho - \, \frac{\Gamma^2_\rho}{4} 
\;\; , \;\; \mbox{Im}^2\! \left[ M_\rho^2 \right] = m_\rho^2 \, \Gamma^2_\rho
\;\; .\end{eqnarray}
For convenience, we can shift the integration variable $s_2$ by
$s_2=:\bar{s}_2+(s-s_1+3\, m^2_\pi)/2$. Then, the factor $\Big[ \,
- s^2_1 \, s_2 + s_1 \, s_2 (s -s_2 + 3\, m^2_\pi ) - m^2_\pi (s-\, m^2_\pi)^2
\; \Big]$ reduces to $s_1 \, ((\bar{s}^{\,+}_2)^2 - \bar{s}^{\,2}_2)$,
with $\bar{s}_2^{\,\pm}:=  \pm \sqrt{\lambda (s,m^2_\pi ,s_1) \, \lambda (s_1 ,
m^2_\pi , m^2_\pi )} /(2\,s_1)$.
Using the new integration variable $\bar{s}_2$, we get for the decay width
the expression 
\begin{eqnarray}
\lefteqn{\Gamma_{\omega\rightarrow\rho\pi\rightarrow \pi\pi\pi} (\vec{p}\,)
 \simeq \frac{|
 \sqrt{2} \, g^{\,\prime} \; G^{\,\prime}_\pi |^2}{2\, \sqrt{s} \; (2\pi)^5} 
\; \frac{\pi^2}{4\, s} \int^{\,(\sqrt{s} - \, m_\pi)^2}_{\,(2\, m_
\pi)^2} \!\! ds_1 \;\; \frac{1}{| s_1 - M_\rho^2 |^2} }
\nonumber \\
 & & \nonumber \\
 & & 
\Bigg\{ \; \frac{\sqrt{\lambda^3 (s,m^2_\pi ,s_1) \, \lambda^3 (s_1 ,
m^2_\pi , m^2_\pi )}}{6\,s^2_1} \;
+  2\,s_1 \, \int^{\, \bar{s}^{\,+}_2}_{\, \bar{s}^{\,-}_2} d\bar{s}_2 
\;\; ((\bar{s}^{\,+}_2)^2 - \bar{s}^{\,2}_2) \;
\nonumber \\
 & & \nonumber \\
 & &   
 \frac{\left(\bar{s}_2+\frac{1}{2} (s-s_1+3\, m^2_\pi) - \mbox{Re} 
\left[ M_\rho^2 \right]\right)\left(s_1 - \mbox{Re} 
\left[ M_\rho^2 \right]\right) + \mbox{Im}^2\! \left[ M_\rho^2 \right]}
{\left(\bar{s}_2+\frac{1}{2}(s-s_1+3\, m^2_\pi) - \mbox{Re} 
\left[ M_\rho^2 \right]\right)^2 + \mbox{Im}^2\! \left[ M_\rho^2 \right]}
 \; \Bigg\} \label{fin1}
\; .\end{eqnarray}

For the $s_1$ integration we have to rewrite the modulus squared of the
$\rho$ propagator in terms of the real and imaginary parts of the
complex mass $M_\rho = m_\rho - i\, \Gamma_\rho /2$, i.e.,
\begin{equation}
 \frac{1}{|\,s_1 - M_\rho^2\,|^2} \, = \, \frac{1}{|\,s_1 - (m_\rho -
i\, \frac{\Gamma_\rho}{2}\,) \,|^2} \, = \,
\frac{1}{(\,s_1 - m^2_\rho +
\frac{\Gamma^2_\rho}{4} )^2 +  m^2_\rho \, \Gamma^2_\rho}
\; .\end{equation}
The two-dimensional integral in (\ref{fin1})
can easily be evaluated numerically, after having set
$\sqrt{s} = m_\omega$ and remembering the identities $\lambda
(s,m^2_\pi ,s_1)  =  (s- \, m^2_\pi - \,
s_1)^2 - 4 \, m^2_\pi\, s_1$ and
$\lambda (s_1 , m^2_\pi , m^2_\pi ) = s_1 \, (s_1 - \, 4 \, m^2_\pi )$.
Substituting $m_\omega=$ 0.782 GeV, $m_\pi=$ 0.140 GeV, $m_\rho=$ 0.770 GeV, 
and $\Gamma_\rho=$ 0.151 GeV, we obtain for the double integral the result
$3.35\times10^{-3}$ (GeV)$^6$. 

Furthermore,
the coupling $g'\simeq g =4.2 \,(\pm 0.1)$ has been determined from
the decay width $\Gamma (\rho^0\rightarrow \pi^+\pi^-) = g^2 q^3_\pi/(3\pi
\, m^2_\rho) = 151.2 \pm 1.2\;\xrm{MeV}$ with $q_\pi = \sqrt{\lambda (m^2_{
\rho^0}, m^2_{\pi^+}, m^2_{\pi^-})/(4\,m^2_{\rho^0})} = \sqrt{(m^2_{
\rho^0} -  4\, m^2_{\pi^\pm})/4}$,
while $G^{\,\prime}_\pi \simeq G_\pi = \sqrt{2}\,g\,g_{\omega\pi^0\gamma}$
follows from the partial decay width
$\Gamma(\omega\rightarrow\pi^0\gamma) \simeq
\alpha_{{}_{\mbox{\scriptsize QED}}}\,g^2_{\omega\pi^0\gamma}\,q_
\gamma^3/3$ = 0.74 MeV, where $q_\gamma = \sqrt{\lambda (m^2_{\omega},
m^2_{\pi^0},0)/(4 \,m^2_{\omega})} = (m_\omega^2-m_{\pi^0}^2)/(2m_\omega)$
= \mbox{0.379 GeV} (see also Ref.~\cite{Bra95,coo1}). 
With $\alpha_{{}_{\mbox{\scriptsize QED}}}\simeq 1/137$, we thus get
\[g_{\omega\pi^0\gamma}=\sqrt{\frac{3\Gamma(\omega\rightarrow\pi^0\gamma)}
{\alpha_{{}_{\mbox{\scriptsize QED}}}\,q^3_\gamma}}=
2.36\mbox{\,\,GeV}^{-1},\]
which gives $G^{\,\prime}_\pi \simeq $ 14.0 GeV$^{-1}$. 
Combining all, we 
finally obtain our prediction
\begin{eqnarray}\Gamma_{\omega\rightarrow\rho\pi\rightarrow \pi\pi\pi}
 & = &6.11 \mbox{\,\,MeV}\; .\nonumber \end{eqnarray}
In order to compare this with experiment, we must first try to determine which
fraction of the total $\omega$ width of 8.44 MeV \cite{PP} is due to the
cascade $\omega\rightarrow \rho\pi\rightarrow \pi\pi\pi$. Now, also according
to the most recent {\sl Review of Particle Properties} \cite{PP}, the $\omega$
ows about 89\% of its width to three-pion decays. However, part of this may be
the result of direct, {\em cascadeless} \/$3\pi$ decays, which can be
understood as OZI decays with the ``simultaneous'' creation of {\em two}
\/$q\bar{q}$ pairs, or as a ``quark-box contribution'' in the language of
Lucio-Martinez {\em et al.}, Ref.~\cite{LNSV}.
 Unfortunately, there is no
experimental information on the proportion of ``direct'' vs.\ (virtual)
``cascade'' decays of the $\omega$. Nevertheless, as already mentioned in the 
introduction, such information does exist
for the $\phi$(1020), according to which the $\rho\pi$ mode accounts for at
least 80\% of the total three-pion decays, provided that interference between
the $\rho\pi$ and (direct) $3\pi$ modes is neglected \cite{PP,Par76}.
Considering that the $3\pi$ decays of the $\phi$(1020) are due to its (small)
non-strange ($n\bar{n}$) component, it seems reasonable to extrapolate this
finding to the $\omega$(782). Thus, we get an estimated experimental width
\[\Gamma_{\omega\rightarrow\rho\pi\rightarrow\pi\pi\pi}^
{\,\mbox{\scriptsize exp}} \geq 0.80 \times 0.89 \times 8.44 \mbox{\,\,MeV}
= 6.0\mbox{\,\,MeV} , \]
with an ``upper bound'' of $0.89\times 8.44 = 7.5$ MeV, which fully agrees with
our UERM prediction.

Also note that our $G^{\,\prime}$ of 14.0 GeV$^{-1}$ is in good agreement
with the experimental data (14$\pm$2 GeV$^{-1}$ \cite{rhompi}),
in particular as compared to the interpolated result for
$g_{\omega\rho\pi}$ in Table I of Ref.~\cite{LNSV} for a
$\Gamma^{\,\mbox{\scriptsize GSW}}(\omega\rightarrow3\pi)$
equal to the value we just found.
\clearpage

\section{Conclusion and discussion}
\label{conclusion}
In the present paper we have extended a formalism, previously applied
to fermions, to bosonic systems, i.e., charged and uncharged
scalar, pseudoscalar and vector mesons. This method allows to handle, in a
field-theoretic framework, resonant degrees of freedom in the intermediate
state on the same footing as stable particles.

A simple yet topical application of the method to the three-pion decay
of the $\omega$ meson, via the cascade process (\ref{cascade}), results
in an $\omega$ partial decay width in full agreement with experiment.

In principle, the formalism can be applied to many other known cascade
processes, not only by effective field-theoretic methods, but also in 
quantum-mechanical coupled-channel approaches.

As already stated above, the application of the UERM to processes
with resonant particles in the initial or final state is a task for future
investigation, which may have a large impact on the understanding of 
how to construct relativistic optical ``potentials''.

As an outlook, we also want to stress here several points that have non-trivial
consequences for standard QFT, as induced by the UERM in the limit of vanishing
width:

The extension of the UERM to massless fermions (like (anti-)neutrinos) is
straightforward and well-defined, if one keeps the width of such degrees of
freedom finite. Yet the so called ``chiral symmetry'' of such systems gets,
within the UERM, a completely new meaning, as it leads to an interchange of
the (anti-)particle and (anti-)hole sector of the theory, which cannot be
resolved in an equivalent and adequate way by standard QFT.
A simple example is the simultaneous existence of a conserved axial current
(CAC) and an inequivalent partially conserved axial current (PCAC) in such
a vector-current-conserving (CVC) theory. 

The UERM treatment of massless bosons is only well-defined, if both the
imaginary {\em and} real part of their self-energies stay at least
infinitesimally non-zero. In case of a vanishing real part of the
self-energy of bosons, the causality properties of (anti-)particles and
(anti-)holes coincide, giving rise to formal pathologies, to be resolved
only by careful limiting procedures.
 
Once this kind of complications are under control, it is a challenging task
to extend the UERM to (massless) gauge bosons, and to investigate how to
introduce and understand abelian and nonabelian local gauge symmetries
within the UERM. After all, the consideration of gauge theories with chiral
fermions within the UERM will lead to a much more well-defined and
unambiguous treatment of the anomaly sector of such chiral gauge theories
than in standard QFT.

\section*{Acknowledgements}
F.K. is thankful for the warm hospitality and excellent working atmosphere at
the CFIF (IST) in Lisbon, and the Physics Department of Coimbra University.
The critical remarks and very stimulating discussions on related topics
with E.\ Akhmedov, A.R.\ Bohm, M.\ Dillig, L. Ferreira, A.I.\ Machavariani,
P.\ Maris, A.A.\ Osipov, L.v.\ Smekal and J.\ Zinn-Justin are also
gratefully acknowledged.\\
This work is partly supported by the
{\it Funda\c{c}\~{a}o para a Ci\^{e}ncia e a Tecnologia}
of the {\it Minist\'{e}rio da
Ci\^{e}ncia e da Tecnologia} \/of Portugal,
under Grant no.\ PRAXIS XXI/\-BPD/\-20186/\-99 and
under contract numbers
PESO/\-P/\-PRO/\-15127/\-99,
POCTI/\-35304/\-FIS/\-2000,
and
CERN/\-P/\-FIS/\-40119/\-2000.

\end{document}